\documentclass[10pt]{article} 
\usepackage{fullpage}
\usepackage{array}
\usepackage{amssymb}
\usepackage{bussproofs}
\usepackage{cmll}
\usepackage{fixltx2e}
\usepackage{url}
\usepackage{proof}
\usepackage{stmaryrd}
\usepackage{graphicx}
\usepackage{xspace}
\usepackage[all]{xy}
\usepackage{listings}
\usepackage{multicol}

\input{Qcircuit}

\newcommand{\omitnow}[1]{}
\newcommand{\boolt}{\textsf{Bool}} 
\newcommand{\bfalse}{\texttt{\textbf{false}}\xspace}
\newcommand{\btrue}{\texttt{\textbf{true}}\xspace}

\newcommand{\alt}{~|~}

\newcommand{\singleton}[1]{\{#1\}}

\newcommand{\ip}[2]{\langle #1 \alt #2 \rangle}

\newcommand{\scalarZ}{\bfalse}
\newcommand{\scalarO}{\btrue}

\newcommand{\scalarPlus}{~\ensuremath{\underline{\vee}}~}
\newcommand{\scalarTimes}{\ensuremath{\wedge}}

\newcommand{\usat}{\texttt{UNIQUE-SAT}}

\begin{document}
\title{Solving \usat\ in a Modal Quantum Theory} 
\author{Jeremiah Willcock and Amr Sabry \\
School of Informatics and Computing \\
Indiana University} 
\maketitle

\begin{abstract}
In recent work, Benjamin Schumacher and Michael~D. Westmoreland investigate a
version of quantum mechanics which they call \emph{modal quantum
theory}. This theory is obtained by instantiating the mathematical framework
of Hilbert spaces with a finite field instead of the field of complex
numbers. This instantiation collapses much the structure of actual quantum
mechanics but retains several of its distinguishing characteristics including
the notions of superposition, interference, and entanglement. Furthermore,
modal quantum theory excludes local hidden variable models, has a no-cloning
theorem, and can express natural counterparts of quantum information
protocols such as superdense coding and teleportation.

We show that the problem of \usat\ --- which decides whether a given Boolean
formula is unsatisfiable or has exactly one satisfying assignment --- is
deterministically solvable in any modal quantum theory \emph{in constant
time}. The solution exploits the lack of orthogonality in modal quantum
theories and is not directly applicable to actual quantum theory.
\end{abstract}

\section{Modal Quantum Theory}
\label{sec:dqt} 

In their recent work, Schumacher and Westmoreland~\cite{modalqm} argue that
much of the structure of traditional quantum mechanics is maintained in the
presence of finite fields. In particular, they establish that the quantum
theory based on the finite field of booleans retains the following
characteristics of quantum mechanics: the notions of superposition,
interference, entanglement, and mixed states of quantum systems; the time
evolution of quantum systems using invertible linear operators; the
complementarity of incompatible observables; the exclusion of local hidden
variable theories and the impossibility of cloning quantum states; and the
presence of natural counterparts of quantum information protocols such as
superdense coding and teleportation.

\paragraph*{Fields.} 
A field is an algebraic structure with notions of addition and multiplication
that satisfy the usual axioms.  The rationals, reals, complex numbers, and
quaternions form fields that are infinite. There are also finite fields that
satisfy the same set of axioms. Finite fields are necessarily ``cyclic.''

The simplest field is the field of booleans $\mathbb{F}_2$ consisting of two
scalars $\{ \scalarZ, \scalarO\}$.  The elements $\scalarZ$ and $\scalarO$
are associated with the probabilities of quantum events, with $\scalarZ$
interpreted as \emph{definitely no} and $\scalarO$ interpreted as
\emph{possibly yes}.\footnote{Everything works if we switch the
interpretation with $\scalarZ$ interpreted as \emph{possibly no} and
$\scalarO$ as \emph{definitely
yes}~\cite{DBLP:journals/corr/abs-0910-2393}. } The field $\mathbb{F}_2$
comes with an addition operation~\scalarPlus\ (which in this case must be
exclusive-or) and a multiplication operation~\scalarTimes\ (which in this
case must be conjunction). In particular we have:

\[\begin{array}{rclcl@{\qquad\qquad\quad}rclcl}
\scalarZ &\scalarPlus& \scalarZ &=& \scalarZ & 
          \scalarZ &\scalarTimes& \scalarZ &=& \scalarZ \\
\scalarZ &\scalarPlus& \scalarO &=& \scalarO &
          \scalarZ &\scalarTimes& \scalarO &=& \scalarZ \\
\scalarO &\scalarPlus& \scalarZ &=& \scalarO &
          \scalarO &\scalarTimes& \scalarZ &=& \scalarZ \\
\scalarO &\scalarPlus& \scalarO &=& \scalarZ &
          \scalarO &\scalarTimes& \scalarO &=& \scalarO 
\end{array}\]
The definitions are intuitively consistent with the interpretation of
scalars as probabilities for quantum events except that it appears
strange to have $\scalarO \scalarPlus \scalarO$ be defined as
$\scalarZ$, i.e., to have a twice-possible event become
impossible. This results from the cyclic property intrinsic to finite
fields requiring the existence of an inverse to $\scalarPlus$. This
inverse must be $\scalarO$ itself which means that $\scalarO$
essentially plays both the roles of ``possible with phase~1'' and
``possible with phase -1'' and that the two occurrences cancel each
other in a superposition.

\paragraph*{Vector Spaces.} 
Consider the simple case of a 1-qubit system with bases $\ket{0}$
and~$\ket{1}$. The Hilbert space framework allows us to construct an infinite
number of states for the qubit all of the form $\alpha\ket{0} + \beta\ket{1}$
with~$\alpha$ and $\beta$ elements of the underlying field of complex numbers
and with the side condition that $|\alpha|^2+|\beta|^2=1$. Moving to a finite
field immediately limits the set of possible states as the coefficients
$\alpha$ and~$\beta$ are now drawn from a finite set. In the field
$\mathbb{F}_2$, there are exactly three valid states for the qubit:
$\scalarZ\ket{0} ~+~ \scalarO\ket{1}$ (which is equivalent to~$\ket{1}$),
$\scalarO\ket{0} ~+~ \scalarZ\ket{1}$ (which is equivalent to $\ket{0}$), and
$\scalarO\ket{0} ~+~ \scalarO\ket{1}$ (which we write as $\ket{+}$).  The
fourth possibility is the zero vector which is not an allowed quantum
state. In a larger field with three scalars, there would be eight possible
states for the qubit which intuitively suggests that one must ``pay'' for the
amount of desired superpositions: the larger the finite field, the more
states are present with the full Bloch sphere seemingly appearing at the
``limit.''

Interestingly, we can easily check that the three possible vectors for a
1-qubit state are linearly dependent with any pair of vectors expressing the
third as a superposition:
\[\begin{array}{rclcl}
\ket{0} &+& \ket{1} &=& \ket{+} \\
\ket{0} &+& \ket{+} &=& \ket{1} \\
\ket{1} &+& \ket{+} &=& \ket{0} 
\end{array}\]
In other words, other than the standard basis consisting of $\ket{0}$ and
$\ket{1}$, there are just two other possible bases for this vector space,
$\singleton{\ket{1}, \ket{+}}$ and $\singleton{\ket{+},\ket{0}}$. 

\paragraph*{Inner Products.}
A Hilbert space comes equipped with an inner product $\ip{v_1}{v_2}$ which is
an operation that associates each pair of vectors with a complex scalar value
that quantifies the ``closeness'' of the two vectors. The inner product
induces a norm $\sqrt{\ip{v}{v}}$ that can be thought of as the length of
vector $v$. In a finite field, we can still define an operation
$\ip{v_1}{v_2}$ which, following our interpretation of the scalars, would
need to return $\scalarZ$ if the vectors are definitely not the same and
$\scalarO$ if the vectors are possibly the same. This operation however does
\emph{not} yield an inner product, as the definition of inner products
requires that the field has characteristic 0. This is not the case for the
field $\mathbb{F}_2$ (nor for any finite field for that matter) as the sum of
positive elements must eventually ``wrap around.'' In other words, if we
choose to instantiate the mathematical framework of Hilbert spaces with a
finite field, we must therefore drop the requirement for inner products and
content ourselves with a plain vector space.

\paragraph*{Invertible Linear Maps.} 
In actual quantum computing, the dynamic evolution of quantum states is
described by unitary maps which preserve inner products. As modal quantum
theory lacks inner products, the dynamic evolution of quantum states is
described by any invertible linear map, i.e., by any linear map that is
guaranteed never to produce the zero vector.

As an example, there are 16 linear (not necessarily invertible) functions in
the space of 1-qubit functions. Out of these, six are permutations on the
three 1-qubit vectors; the remaining ten maps all map one of the vectors to
the zero vector which makes them non-invertible. On one hand, this space is
quite impoverished compared to the full set of 1-qubit linear maps in the
Hilbert space. In particular, even some of the elementary unitary maps such
as the Hadamard matrix are not expressible in that space. On the other hand,
the space includes non-unitary maps that are not allowed in actual quantum
computing. Of particular interest are the following two maps:
\[\begin{array}{rcl@{\qquad\qquad\qquad}rcl}
s~\ket{0} &=& \ket{+} & s^\dagger~\ket{0} &=& \ket{0} \\
s~\ket{1} &=& \ket{1} & s^\dagger~\ket{1} &=& \ket{+} \\
s~\ket{+} &=& \ket{0} & s^\dagger~\ket{+} &=& \ket{1} 
\end{array}\]
The space of 1-qubit maps also includes the identity map which we refer to as
$X_0$ below, and the negation map which we refer to as $X_1$ below.

\paragraph*{Measurement.} 
Measurement in arbitrary bases is complicated but measurement in the standard
basis is fairly simple. In a 1-qubit system, we have:
\begin{itemize}
\item $\textit{measure}~\ket{0}$ deterministically produces $\ket{0}$;
\item $\textit{measure}~\ket{1}$ deterministically produces $\ket{1}$;
\item $\textit{measure}~\ket{+}$ non-deterministically produces $\ket{0}$ or
$\ket{1}$.
\end{itemize}
There is no probability distribution associated with the non-deterministic
choice between $\ket{0}$ and $\ket{1}$ in the last case.

\paragraph*{Entanglement and Superdense Coding.} 
Despite the restriction to finite fields and the drastic reduction in the
state space of qubits and their maps, the theory built on the field of
booleans has a definite quantum character: it can, for example, express
quantum protocols such as superdense coding~\cite{modalqm}.

\section{\usat} 

The problem of \usat\ is the problem of deciding whether a given Boolean
formula known to have either 0 or 1 satisfying assignment has 0 or 1
assignment. Surprisingly this problem is, in a precise
sense~\cite{Valiant198685}, just as hard as the general satisfiability
problem and hence all problems in the NP complexity class.

\section{Solving \usat\ in the Field of Booleans} 

We are given a classical function $f : \boolt^n \rightarrow \boolt$ that
takes $n$-bits and returns at most one \btrue result. In the following, we
use $\overline{x}$ to denote a sequence $x_1, x_2, \ldots, x_n$ of $n$
bits. Given a function $f : \boolt^n \rightarrow \boolt$, we construct the
Deutsch quantum black box $U_f$ as follows:
\[
  U_f \ket{y}\ket{\overline{x}} ~=~
    \ket{y \scalarPlus f(\overline{x}}\ket{\overline{x}}
\]
It is straightforward to verify that $U_f$ is an invertible map, and hence an
acceptable map for the evolution of states in a modal quantum theory.

\paragraph*{The Algorithm.} 
 
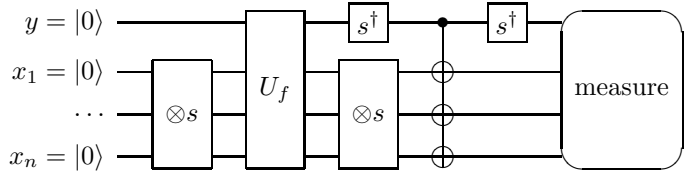
\begin{figure}[t]
\[\begin{array}{c}
\Qcircuit @C=1.3em @R=.7em {
\lstick{y=\ket{0}}   & \qw                      & \multigate{3}{U_f} &  \gate{s^\dagger}         & \ctrl{3} & \gate{s^\dagger} & \multimeasure{3}{\text{measure}} \\
\lstick{x_1=\ket{0}} & \multigate{2}{\otimes s} & \ghost{U_f}        &  \multigate{2}{\otimes s} & \targ    & \qw              & \ghost{\text{measure}}           \\
\lstick{\ldots}      & \ghost{\otimes s}        & \ghost{U_f}        &  \ghost{\otimes s}        & \targ    & \qw              & \ghost{\text{measure}}           \\
\lstick{x_n=\ket{0}} & \ghost{\otimes s}        & \ghost{U_f}        &  \ghost{\otimes s}        & \targ    & \qw              & \ghost{\text{measure}}           
}
\end{array}\]
\caption{\label{fig:alg}Modal Quantum Circuit for \usat in $\mathbb{F}_2$}
\end{figure}

The algorithm is pictorially presented in Figure~\ref{fig:alg}. It consists
of the following steps:
\begin{enumerate}
\item Initialize an $n+1$ qubit state to $\ket{0}\ket{\overline{0}}$.
\item Apply the map $s$ to each qubit in the second component of
the state.
\item Apply the quantum black box version $U_f$ to the entire state. 
\item Apply the map $s$ to each qubit in the second component of the state.
\item Apply the map $s^\dagger$ to the first component of the state.
\item Let the first component of the state be $a$. Apply the map $X_a$ to
   each qubit in the second component of the state.
\item Apply the map $s^\dagger$ to the first component of the state.
\item Measure the resulting state in the standard basis for $n+1$ qubits. If
the measurement is $\ket{0}\ket{\overline{0}}$ then the function $f$ is
\textbf{unsatisfiable}. If the measurement is anything else then the
function $f$ \textbf{satisfiable}.
\end{enumerate} 

\paragraph*{Correctness (case I).} Assume the function $f$ is unsatisfiable:
\begin{enumerate} 
\item In the first step, the initial state is $\ket{0}\ket{\overline{0}}$
\item In the second step, the state becomes $\ket{0}\ket{\overline{+}}$.
\item In the third step, the function $U_f$ is the identity and the state
remains $\ket{0}\ket{\overline{+}}$.
\item Applying $s$ to each qubit in the second component of the state
produces $\ket{0}\ket{\overline{0}}$.
\item Applying $s^\dagger$ to the first component leaves the state unchanged
as $\ket{0}\ket{\overline{0}}$.
\item As the first component of the state is 0, applying the map $X_0$ (which
is the identity) leaves the state unchanged as $\ket{0}\ket{\overline{0}}$.
\item Applying $s^\dagger$ to the first component leaves the state unchanged
as $\ket{0}\ket{\overline{0}}$.
\item Measuring the state will deterministically produce
$\ket{0}\ket{\overline{0}}$.
\end{enumerate} 

\paragraph*{Correctness (case II).} Assume the function $f$ is satisfiable 
at some input $a_1, a_2, \ldots, a_n$ denoted $\overline{a}$:
\begin{enumerate} 
\item In the first step, the initial state is $\ket{0}\ket{\overline{0}}$
\item In the second step, the state becomes $\ket{0}\ket{\overline{+}}$. We
can write this state as $\ket{0}\ket{\overline{a}} + \Sigma_{\overline{x}}^*
\ket{0}\ket{\overline{x}}$ where the superscript ${}^*$ is to remind us that
the summation is over all the $2^n$ combinations except $a_1, a_2, \ldots,
a_n$.
\item In the third step, applying $U_f$ produces $\ket{1}\ket{\overline{a}}
+ \Sigma_{\overline{x}}^* \ket{0}\ket{\overline{x}}$. Because
$\ket{0}\ket{\overline{a}} + \ket{0}\ket{\overline{a}}$ is the zero vector, we can
rewrite this state as $\ket{+}\ket{\overline{a}} + \Sigma_{\overline{x}}
\ket{0}\ket{\overline{x}}$ where the summation is now over the all vectors, i.e.,
we can write the state as $\ket{+}\ket{\overline{a}} +
\ket{0}\ket{\overline{+}}$.
\item Applying $s$ to each qubit in the second component produces 
$\ket{+}\ket{\overline{s(a)}} + \ket{0}\ket{\overline{0}}$.
\item Applying $s^\dagger$ to the first component produces: 
$\ket{1}\ket{\overline{s(a)}} + \ket{0}\ket{\overline{0}}$.
\item Applying $X_b$ where $b$ is the first component of the state to each
qubit in the second component to the state produces
$\ket{1}\ket{\overline{\textit{not}(s(a))}} + \ket{0}\ket{\overline{0}}$.
\item Applying $s^\dagger$ to the first component produces 
$\ket{+}\ket{\overline{\textit{not}(s(a))}} + \ket{0}\ket{\overline{0}}$.
\item For the measurement of $\ket{+}\ket{\overline{\textit{not}(s(a))}} +
\ket{0}\ket{\overline{0}}$ to be guaranteed to be never
$\ket{0}\ket{\overline{0}}$, we need to verify that
$\ket{+}\ket{\overline{\textit{not}(s(a))}}$ has one occurrence
$\ket{0}\ket{\overline{0}}$. This can be easily proved as follows. Since each
$a_i$ is either 0 or 1, then each $s(a_i)$ is either $+$ or $1$, and hence
each $\textit{not}(s(a_i))$ is either $+$ or $0$. The result follows since
any state with a combination of $+$ and $0$ --- when expressed in the standard
basis --- would consist of a superposition containing the state
$\ket{0\ldots}$.
\end{enumerate} 

\section{Conclusion} 

We have presented an algorithm for solving \usat\ in the modal quantum theory
based over the finite field of booleans. The existence of this algorithm
suggests that modal quantum theories, as defined, collapse too much of the
structure of quantum mechanics and that they should be restricted to retain
some notion of orthogonality.

\section*{Acknowledgments}
We have benefited from many discussions with the Quantum and Natural
Computing group at Indiana University including Jerry Busemeyer, Mike Dunn,
Andy Hanson, Andrew Lumsdaine, Larry Moss, and Gerardo Ortiz.

\bibliographystyle{plain}
\bibliography{p}
\end{document}